\documentclass[pre,reprint,amsmath,amssymb,aps,longbibliography]{revtex4-2}
\usepackage{amsmath}
\usepackage{graphicx}
\usepackage{xcolor}
\usepackage{upgreek}
\usepackage{dcolumn}
\usepackage{bm}
\usepackage{color}
\usepackage{mathtools}
\usepackage[normalem]{ulem}
\usepackage[english]{babel}
\usepackage{xr-hyper}
\usepackage[colorlinks=true,linkcolor=blue,urlcolor=blue,citecolor=blue]{hyperref}

\newcommand{\fig}[1]{Fig.~\ref{#1}}

\newcommand{\Fig}[1]{Figure~\ref{#1}}
\newcommand{\eq}[1]{Eq.~(\ref{#1})}

\newcommand{\dma}{\mathrm{d}}

\newcommand{\roundbk}[1]{\left({#1}\right)}
\newcommand{\squarebk}[1]{\left[{#1}\right]}

\newcommand{\dE}{\Delta E}

\begin{document}
\title{Surface mobility gradient and emergent facilitation in glassy films}
\author{Qiang Zhai$^1$} 
\author{Xin-Yuan Gao$^2$}
\author{Chun-Shing Lee$^3$}
\author{Chin-Yuan Ong$^4$}
\author{Ke Yan$^5$}
\email{ yanke@mail.xjtu.edu.cn}
\author{Hai-Yao Deng$^6$}
\email{dengh4@cardiff.ac.uk}
\author{Sen Yang$^1$}
\email{yangsen@xjtu.edu.cn}
\author{Chi-Hang Lam$^7$}
\email{C.H.Lam@polyu.edu.hk}

\address{
$^1$School of Physics, MOE Key Laboratory for Nonequilibrium Synthesis and Modulation of Condensed Matter, Xi'an, Shaanxi, 710049, China\\
$^2$Department of Physics, The Chinese University of Hong Kong, Shatin, New Territories, Hong Kong, China\\
$^3$School of Science, Harbin Institute of Technology (Shenzhen), Shenzhen, 518055, China \\
$^4$School of Physics, Yale University, New Haven, Connecticut, 06520, USA\\
$^5$School of Mechanical Engineering, Xi'an Jiaotong University, Xi'an, Shaanxi,710049,China\\
$^6$School of Physics and Astronomy, Cardiff University, 5 The Parade, Cardiff CF24 3AA, Wales, UK \\
$^7$Department of Applied Physics, Hong Kong Polytechnic University, Hong Kong, China 
}
	
\date{\today}

\begin{abstract}
Confining glassy polymer into films can substantially modify  their local and film-averaged properties. We present a lattice model of film geometry with void-mediated facilitation behaviors but free from any elasticity effect.  We analyze the spatially varying viscosity to delineate the transport property of glassy films. The film mobility measurements reported by  [Yang et. al., Science, 2010, 328, 1676] are successfully reproduced. The flow exhibits a crossover  from simple viscous flow to a surface-dominated regime as temperature decreases. The propagation of a highly mobile front induced by the free surface is visualized in real space. Our approach provides a microscopic treatment of the observed glassy phenomena.
\end{abstract}

\maketitle	

\section{Introduction}

{Glassy films have drawn substantial research interests ~\cite{mckenna2017,roth2021review,schweizer:2019}.
The presence of interfaces  either with air or a substrate was reported to alter the glass transition temperature $T_\mathrm{g}$~\cite{keddie1994,forrest1996,pye2011,baumchen2012,inoue2018}, film viscosity~\cite{yang2010, evans2012,forrest2014} and relaxation  time~\cite{frieberg2012,sun2022}.  In general,  a free surface tends to facilitate the relaxation while a strongly attractive substrate or adsorbed layer tends to slow down the dynamics~\cite{torres2000}. Early studies suggest that interfaces induce a gradient  in $T_\mathrm{g}$~\cite{ellison2003}
and the $\alpha$-relaxation time~\cite{varnik2002pre}, both being important characteristics of structural relaxation in glasses. Besides, the observed dynamical gradients penetrate much deeper than the particle density gradients~\cite{varnik2002pre,lam2018film}.
Impacts on confinement effects by polymer chain connectivity~\cite{roth2006} and possible non-equilibrium nature of film samples~\cite{pana2017} are also interesting questions.}

{The reduction of $T_\mathrm{g}$ has been interpreted in various theoretical postulations ~\cite{keddie1994,herminhaus2002,degennes2000,forrest2015string,white2021,mirigian2015,ghan2023}. In particular, a two-layer model, namely a highly mobile non-glassy layer atop of a bulk-like glassy inner layer, is  used to analyze the film mobility inferred from surface roughing experiments~\cite{yang2010}. The two-layer model serves as a simplified approach to account for  a smooth gradient of $T_\mathrm{g}$ and relaxation time~\cite{evans2013,inoue2011,kanaya2015,koga2011,lam2018b}.

{Lattice models have been instrumental in understanding glassy phenomena~\cite{garrahan2011review}. Compared with molecular dynamics (MD) simulations, lattice models are more simplified and aim to capture the essential features of  glassy systems.  Such simplicity often allows detailed analysis and provides insights into the glass transition. Recently, the dynamical facilitation picture of glass \cite{chandler2010review}, originally motivated by lattice models~\cite{fredrickson1984,palmer1984}, has received a surge of interest \cite{scalliet2022,hasyim2023,tahaei2023}, with much efforts focusing on long-range elasticity as a possible mediator of facilitation \cite{tahaei2023,hasyim2023}. 

Here we study dynamics of glassy films using a lattice model with emergent facilitation but free from long-range elasticity. We reproduce a spatial gradient in the relaxation time, which lends support to the two-layer model. We also  demonstrate a void-mediated facilitation mechanism. Our study is based on a distinguishable-particle lattice model (DPLM)~\cite{zhang2017} free from long-range elastic interactions. It has been shown capable of reproducing a plethora of glassy phenomena with little modifications of the model, including Kovacs paradox~\cite{lulli2020} and effect~\cite{lulli2021}, fragility~\cite{lee2020}, heat capacity overshoot \cite{lee2021}, two-level systems \cite{gao2022}, diffusion coefficient power-laws \cite{gopinath2022} and Kauzmann's paradox \cite{gao2023}.  Dynamical facilitation behaviors resembling those from MD simulations are also observed, without invoking elasticity, facilitation rule, or kinetic constraint. 

\section{Film Lattice Model}
\begin{figure}[!htb]
 \begin{centering}
  \includegraphics[width=1\columnwidth]{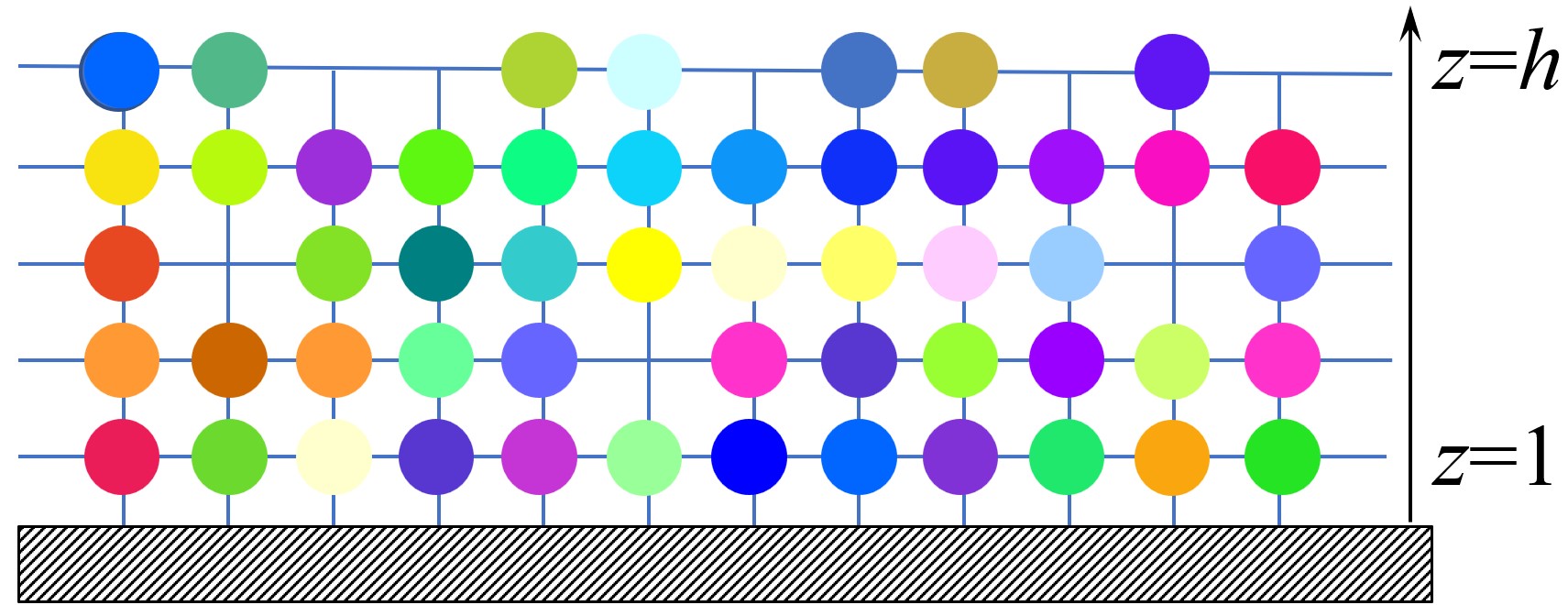}
 \caption{A schematic drawing of the $2D$ lattice model for a supported film of thickness $h$ , in which all the particles are distinguishable.}
 \label{fig:lattice}
\end{centering}
\end{figure}
\begin{figure}[!hbt]
\begin{centering}
\includegraphics[width=0.80\columnwidth]{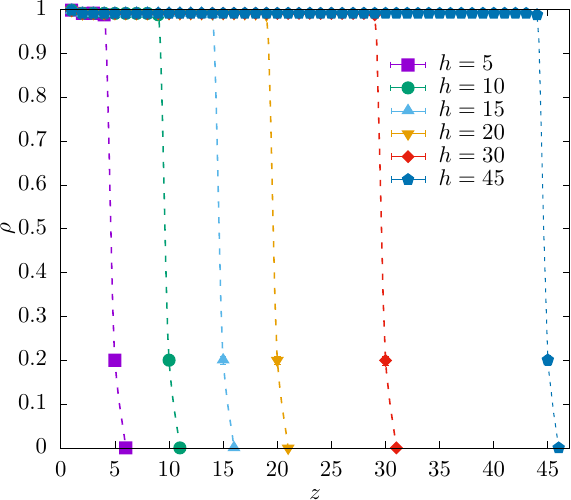}
\caption{The layer-resolved particle density obtained after sample equilibration at $T=0.22$  and thickness $h$. The particle density {in the inner region ($2 \le z \le h-2$)} is nearly constant with a sharp drop at the surface.}
\label{fig:density}
\end{centering}
\end{figure}
\begin{figure}[!hbt]
 \begin{centering}
  \includegraphics[width=0.88\columnwidth]{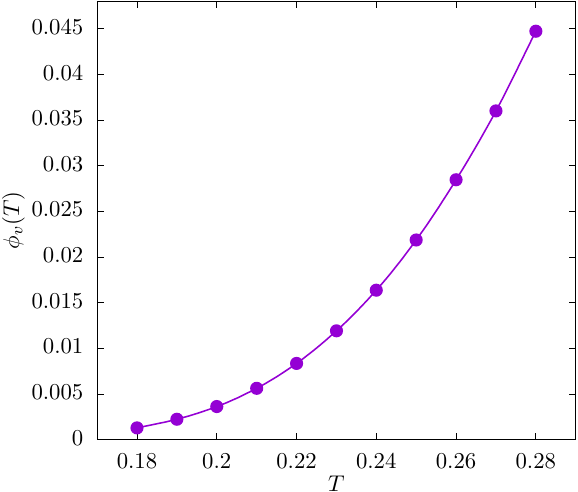}
 \caption{Plot of the void density in the inner region of the film {away from the free surface and the substrate} as a function of $T$.}
\label{fig:phivb}
\end{centering}
\end{figure}

A schematic drawing of the lattice model is displayed in~\fig{fig:lattice}. 
A DPLM film of thickness $h$ in two dimensions (2D) comprises $N$ distinguishable particles -- each of its own type -- sitting on a square lattice terminated along the $z-$axis but with periodic boundary conditions applied along the $x$-axis.
We have recently shown that DPLM behaves qualitatively similarly in two and three dimensions \cite{li2024}.
We let the film be supported on a substrate on one termination (bottom layer, designated by $z=1$) while the other termination is a free surface (top layer, $z=h$). The energy of the film is given by
\begin{equation}
E=\sum_{<i,j>}V_{s_is_j}n_in_j+\epsilon_\mathrm{top}\sum_{i: z_i = h}n_i+\epsilon_\mathrm{bot}\sum_{i: z_i = 1}n_i~, \label{E}
\end{equation}
Here the first summation is over pairs of adjacent sites as indicated by $<i,j>$, $n_i$ is the particle occupation at site $i$, which equals either one if a particle is present or zero if a void (the absence of particle) is present, $V_{s_is_j}$ represents the interaction between a particle at site $i$ of type $s_i$ and a particle at an adjacent site $j$ of type $s_j$, {$\epsilon_\mathrm{top}=1.124$} accounts for the extra energy of a particle at the free surface while $\epsilon_\mathrm{bot}=-0.5$ accounts for substrate effects, and $z_i=1,2,...,h$ is the $z$-coordinate of site $i$. In our system, there are in total $N$ particle types so $s_i = 1,2,...,N$.
The interaction $V_{s_is_j}$ is sampled before the commencement of the simulation from the \emph{a priori} probability distribution, 
\begin{equation}
g(V)=G_0/\roundbk{V_1-V_0}+(1-G_0)\delta(V-V_1),
\end{equation} 
where $V_0=-0.5$, $V_1=0.5$ and $G_0=0.7$.

The equilibrium void-induced dynamics~\cite{yip2020} is simulated by the Metropolis algorithm, with the acceptance rate,
\begin{equation}
w=w_{0} \exp {\squarebk{-\dE \Theta(\dE)}/k_BT},
\end{equation}
where $\dE$ is the change in the system energy $E$ defined by \eq{E} after a particle {hop attempt}. $\Theta(x)=1$ if $x>0$, or $\Theta(x)=0$ otherwise.  The attempt frequency $w_{0} = 10^{6}$  and the Boltzmann constant $k_B$ is set to 1 for simplicity.  

For isothermal equilibrium studies,  {the sample is prepared at the working temperature $T$. Then, a swap algorithm  (swapping among all the particles and voids~\cite{gopinath2022}) is used to speed up the equilibrium process. } The equilibration process continues until the energy and void density are stabilized during more than $10^{10}$ extra swap Monte Carlo attempts.
 
{The particle density measured from equilibrium simulations is displayed in~\fig{fig:density}. Across the temperature range we have studied, the surface void density $\phi_v^s$ is kept at 80\% by fine tuning $N$. This also determines the void density $\phi_v$ ($\ll \phi_v^s$) in the bulk-like inner region which becomes temperature dependent but film thickness independent, as shown in~\fig{fig:phivb}. Only a single topmost layer at $z=h$ has a significantly reduced particle density, modeling a very sharp interface with a high local particle mobility analogous to a free surface. The particle densities are nearly a constant independent of $z$ {except very close to the free surface and the substrate}. Yet, the gradient of the relaxation time covers a thicker region than the density gradient. }

\section{Results and Discussions}
\begin{figure}[!htb]
 \begin{centering}
  \includegraphics[width=0.90\columnwidth]{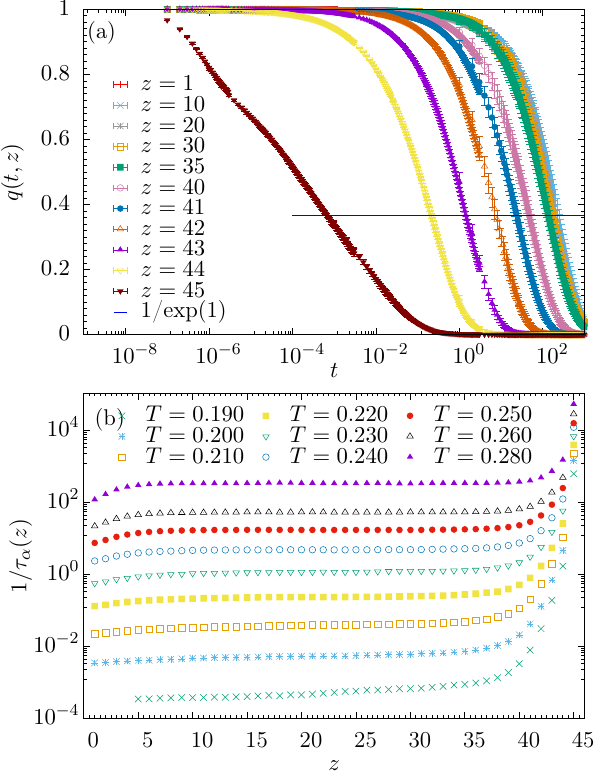}
  \caption{ (a) Layer-resolved overlap function $q(t,z)$ for  film thickness $h=45$ at $T=0.20$. (b) Local relaxation rate $1/\tau_\alpha(z)$ versus distance $z$ from the substrate at various temperatures.}
\label{fig:tau}
\end{centering}
\end{figure}
\begin{figure}[!h]
 \begin{centering}
  \includegraphics[width=0.90\columnwidth]{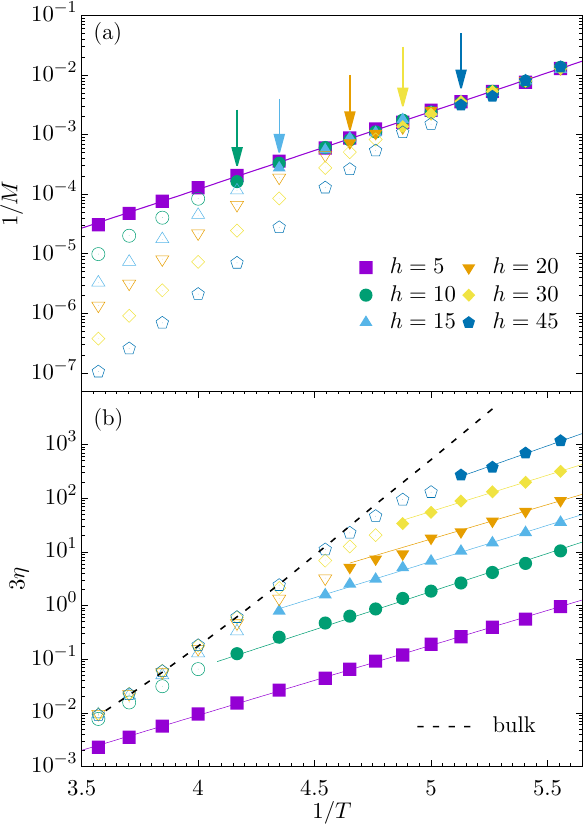}
  \caption{(a) Film mobility $M$ and (b) effective viscosity $\eta$ versus inverse temperature $1/T$ for films of various thickness $h$. Solid and open symbols indicate  surface and bulk-like flow respectively. Arrows show the crossover points between the flow regimes. }
\label{fig:mobility}
\end{centering}
\end{figure}
The depth-resolved $\alpha$-relaxation time $\tau_\alpha(z)$  is extracted from an overlap function $q(t,z)$  obtained under thermal equilibrium.
{The overlap function $q(t,z)$ is defined as the probability that a particle residing initially in layer $z$ experiences no net movement after duration $t$. It has been successfully used to quantify the structural relaxation in  bulk DPLM~\cite{lulli2020}.} In Fig.~\ref{fig:tau}(a), $q(t,z)$ is shown at various $z$ for a film of thickness $h=45$ at a low temperature $T=0.20$. The time window spans over ten orders.  From its value of unity at $t=0$, $q(t,z)$ decays as $t$ increases due to particle motions.  It is seen that for the top layer ($z=45$) the decay is fast but it becomes slow for deeper layers. Then, $\tau_\alpha(z)$ is defined as the time when $q(t,z)$ drops to $1/e$ with $e$ being the Euler constant, i.e. $q(\tau_{\alpha}(z),z) = 1/e$. The extracted local relaxation rate, $1/\tau_\alpha(z)$, is shown at various temperatures in Fig.~\ref{fig:tau} (b). Here one notes that $1/\tau_\alpha(z)$ decays toward a bulk value as $z$ moves into deeper layers; The thickness of the surface affected region well exceeds ten layers {at low $T$}.

Now we employ the as-obtained $\tau_\alpha(z)$ to analyze the film mobility $M$. Our strategy is based on the notion that the local viscosity $\eta(z)$ of a film can be approximated from $\tau_\alpha(z)$, i.e. $\eta(z) = G\tau_\alpha(z)$, known as Maxwell's relation. Here $G$ is the shear modulus of the film, which is assumed with negligible temperature dependence. The mobility $M$ measures the amount of flux that can be carried by the film under a unit pressure gradient parallel to the film~\cite{yang2010}. 

Specifically, we apply the Navier-Stokes equation and the lubrication approximation to film flow along the $x$-direction under no-slip boundary condition at the bottom  and free boundary condition at the top, 
\begin{equation}
\begin{aligned}
\partial_z\squarebk{\eta \roundbk{z} \partial_z {v}_x\roundbk{z}}-\nabla P &= 0~,\\
v_x(z=0)&=0~,\\
\partial_z\squarebk{v_x\roundbk{z}}\vert_{z=h}&=0~,
\end{aligned}
\label{eq:hydro}
\end{equation}
where $P$ is the pressure, $\eta(z)$ is the layer-dependent local viscosity and {$v_x(z)$} is the velocity in the $x$ direction. After solving~\eq{eq:hydro}, we find,
\begin{equation}
v_x(z) = -\nabla P \int_0^z dz' \frac{(h-z')}{\eta(z')}~.
\end{equation}
The total flux is,
\begin{equation}
\begin{aligned}
J&=\int_0^h \mathrm{d} z v_x(z)&=-\nabla P \int_0^h \dma z \int_0^z \dma z' \frac{h-z'}{\eta(z')}~.\\
\end{aligned}
\label{eq:J}
\end{equation}
The mobility of the film is defined as~\cite{yang2010},
\begin{figure*}[!th]
 \begin{centering}
  \includegraphics[width=2\columnwidth]{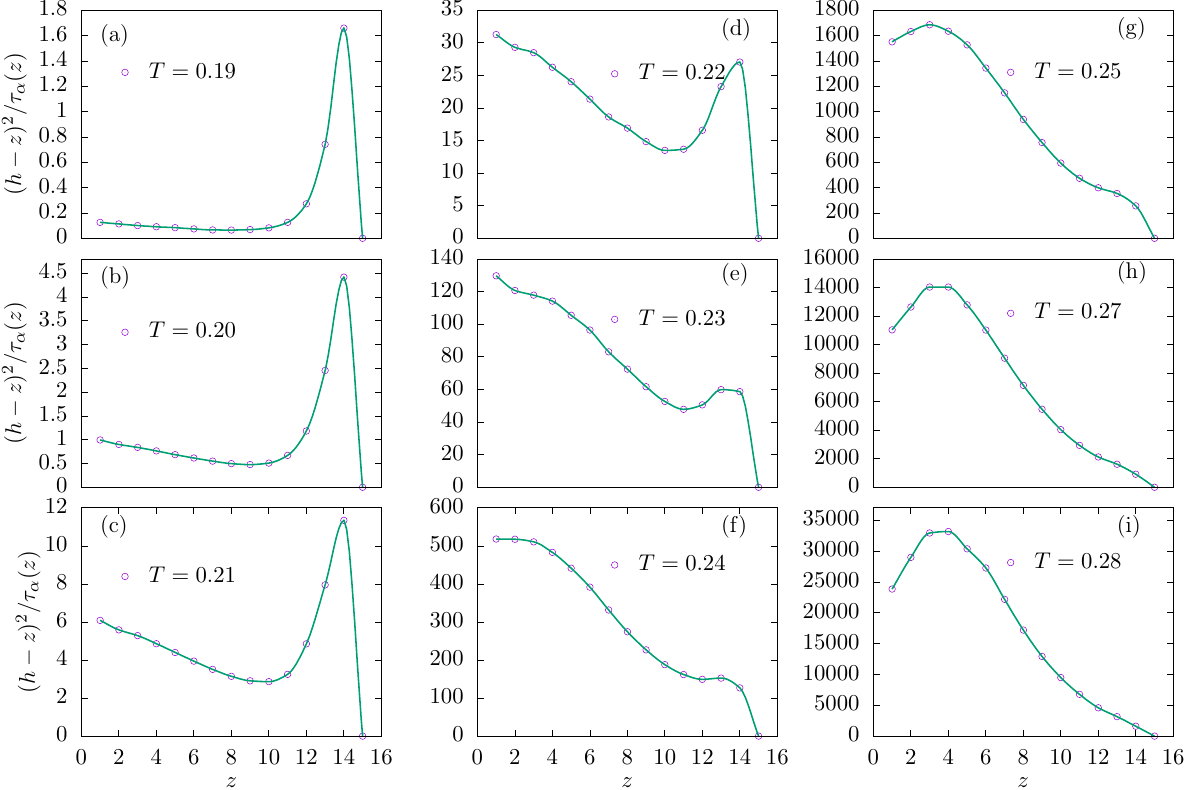}
 \caption{The integrand in~\eq{eq:mobility} is plotted for $h=15$ film at different temperatures. Importance of surface relaxation increases dramatically at $T \le 0.22$, signifying surface transport.}\label{fig:mpeak}
\end{centering}
\end{figure*} 
\begin{equation}
M=-J/\nabla P~.
\label{eq:M}
\end{equation}
We substitute \eq{eq:J} into \eq{eq:M} and integrate by part, leading to
\begin{equation}
\begin{aligned}
M&=\int_{0}^h  \mathrm{d}z\frac{\roundbk{h-z}^2} {\eta(z)}~.
\label{eq:mobility0}
\end{aligned}
\end{equation}
Finally, by approximating $\eta(z)$ with $G\tau_\alpha(z)$ and taking a shear modulus G=1 for simplicity, we reach,
\begin{equation}
M=\int_{0}^h  \mathrm{d}z\frac{\roundbk{h-z}^2} {\tau_\alpha(z)}~.
\label{eq:mobility}
\end{equation}
The effective film viscosity $\eta$ is then defined by $M = \eta^{-1} \int_{0}^h  \mathrm{d}z {\roundbk{h-z}^2}$, which restores $3\eta = h^3/M$ for open-channel viscous flow. In Fig.~\ref{fig:mobility} is shown $M$ and $\eta$ against temperature for films of various thickness. As seen in panel (a), at low temperatures the mobility $M$ of thinner films collapses, indicating the existence of a surface layer dominating the film transport capacity.
The thickness of such a layer is at most five.
In contrast, from panel (b), the effective viscosity $\eta$ of thicker films converges to the bulk value at high temperatures, signifying bulk-like whole-film flow.  All of these features well agree with experimental observations on short-chain polystyrene films~\cite{yang2010}. 

{To better understand the contribution of the local relaxation time $\tau_\alpha(z)$ to the mobility $M$, we plot the integrand,$(h-z)^2/\tau_\alpha(z)$, in~\fig{fig:mpeak}. At high $T$,  the relaxation time varies spatially only by less than two orders of magnitude, and the geometric factor $(h-z)^2$ dominates. In contrast, at low $T$, the spatial variation of the relaxation time becomes much more pronounced  and consequently surface flow dominates the total flow as shown~\fig{fig:mpeak}}.

\begin{figure}[!h]
 \begin{centering}
  \includegraphics[width=1\columnwidth]{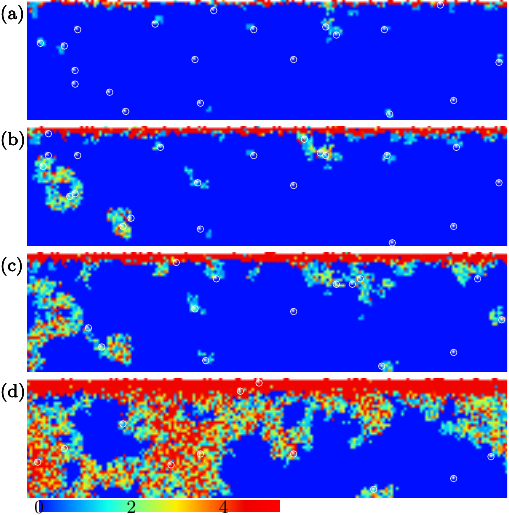}
 \caption{Snapshots showing the temporal evolution of particle displacement  $|r_i(t)-r_i(0)|$ for  film thickness $h=45$ at $T=0.19$ and time $t=2$ (a), $20$ (b), $72$ (c) and  $ \tau_\alpha^\mathrm{bulk}=1550$ (d). White squares, highlighted by circles, denote voids. }
\label{fig:dp}
\end{centering}
\end{figure}
  We now illustrate particle and void dynamics in real space. \Fig{fig:dp} shows snapshots of a film at $T=0.19$ at time $t$ with particles colored according to their displacement  $|\mathbf{r}_i(t)-\mathbf{r}_i(0)|$ where $\mathbf{r}_i(t)$ denotes position of particle $i$ at time $t$. As seen, mobile domains (light blue to red) gradually expand and invade unrelaxed regions (deep blue), demonstrating dynamical facilitation resembling  observations in MD simulations \cite{chandler2011,scalliet2022,herrero2023}.  In the deeper bulk-like inner region, mobile domains vary greatly in sizes, reproducing dynamical heterogeneity characteristic of glass \cite{berthier2011review}.
Facilitation among particles in DPLM are void mediated.
  While smaller domains are induced by single voids, the large ones are initiated by multiple coupled voids which  facilitate each other's motions \cite{zhang2017}.

Moreover, the free surface initiates a rather smooth  mobile front (red) in \fig{fig:dp}.
The large particle displacements there can easily be seen from similar snapshots as  resulting from a continuous flux of isolated voids randomly emerging from and vanishing into the free surface, in contrast to their bulk counterparts which are mostly trapped.
  Motions of voids in DPLM are sub-diffusive at low temperature because of the random energy landscape quenched in the configuration space  \cite{lam2018tree,deng2019}.
This may explain the subdiffusive growth of the mobile front, as observed in MD simulations \cite{herrero2023}. 

Finally, we examine detailed structures of the mobile front in \fig{fig:dp}. The smooth main front discussed above is indeed punctuated by isolated plumes, resembling similar structures reported in previous MD simulations \cite{lam2018b}.
These plumes are often seen to be induced by motions of coupled pairs of voids, as opposed to isolated voids. Despite fewer in number, coupled voids are significantly more mobile due to facilitation \cite{zhang2017} and can penetrate deeper beneath the surface.

\section{Conclusion}
{DPLM encodes the rugged energy landscape typical of glasses through particle-type-dependent interactions. Comparing with the spin glasses~\cite{mezard1990book}, where the disordered interaction is quenched in real space, the interaction in DPLM  is quenched in the configuration space. The void-induced dynamics has recently been evidenced  in a colloidal glass former~\cite{yip2020}.}

To apply findings on films to better understand glass in general as originally envisioned \cite{keddie1994}, a unified approach for both film and bulk geometries is highly desirable. Existing analytical theories on film dynamics \cite{long2001,salez2015,white2021,mirigian2015,ghan2023} inevitably require film-specific assumptions, e.g. a layer-by-layer facilitation-like process in \cite{ghan2023}, and this complicates thorough testing of underlying concepts against both film and bulk phenomena. To this end, lattice models can provide an intermediate bridge between theories and MD simulations or experiments so that assumptions can be reliably tested.  In this work, the DPLM algorithm adopted is identical to that in previous works  on bulk  \cite{lee2020,gao2022,lee2021} with trivial differences only in the boundary layers (see \eq{E}). It provides a truly unified approach with implications generally applicable to both film and bulk.

In conclusion, lattice model simulations in film geometry successfully reproduced  confinement effects characteristic of glassy short-chain polymer or small-molecule films, namely the crossover from whole-film flow to surface flow and {the reduced relaxation time} deep into the film.  Enhanced dynamics as facilitated by the free surface is visualized. Our results show that void-induced particle dynamics is able to account for these phenomena, without invoking long-range elasticity, sample non-equilibrium conditions or polymer chain connectivity. 

\acknowledgments
{This work was supported by China Postdoc Fund Grant No. 2022M722548, Shaanxi NSF Grant No. 2023-JC-QN-0018, Central University Basic Research Fund of China Grant No. xzy012023044,  Hong Kong GRF Grant No. 15303220, National Key Research and Development Program No. 2020YFB2007901 and No. 2022YFE0109500. }

\bibliography{glass_short} 
\end{document}